\documentstyle[aas2pp4]{article}

\tighten
\received{2000 Febrary 2}

\def\orb{{\rm orb}}

 \slugcomment{\it To Appear in ApJ Letters.}

\lefthead{Wood, Montgomery, \& Simpson}
\righthead{SPH Simulations of Nodal Superhumps}

\def\up#1{\raise0.5ex\hbox{#1}}

\begin{document}

\title{Smoothed Particle Hydrodynamics Simulations of Apsidal and
Nodal Superhumps}

\author{Matt A. Wood, Michele M. Montgomery}
\affil{Department of Physics and Space Sciences and SARA Observatory,\\
        Florida Institute of Technology, \\
	Melbourne, FL\ \ 32901-6975, USA,\\
	 wood@astro.fit.edu, michele@astro.fit.edu}

\and

\author{James C. Simpson}
\affil{Computer Sciences Raytheon, P.O. Box 4127, CSR 1310, 
	Patrick Air Force Base, FL 32925-0127, 
	james.simpson@patrick.af.mil}

\begin{abstract}

In recent years a handful of systems
have been observed to show ``negative'' (nodal) superhumps, with
periods slightly {\it shorter} than the orbital period.  It has been
suggested that these modes are a consequence of the slow retrograde
precession of the line of nodes in a disk tilted with respect to the
orbital plane.   Our simulations confirm and refine this model: they
suggest a roughly axisymmetric, retrogradely-precessing, tilted disk
that is driven at a period slightly less than {\it half} the orbital period
as the tidal field of the orbiting secondary encounters in turn the
two halves of the disk above and below the midplane.  Each of these
passings leads to viscous dissipation on one face of an
optically-thick disk --- observers on opposite sides of the disk would
each observe one brightening per orbit, but $180^\circ$ out of phase
with each other.  

\end{abstract}

\keywords{accretion disks
-- binaries: close 
-- cataclysmic variables
-- methods: numerical
-- hydrodynamics
}

\vspace{1in}
\centerline{\large To appear in \it The Astrophysical Journal Letters}

\section{Introduction}

Cataclysmic variable (CV) systems in the mass-ratio range $q\equiv
M_2/M_1\lesssim0.33$ (e.g., Murray 2000) are observed to show the disk 
oscillations known
as superhumps (see Warner 1995 and Patterson 1999 for recent
reviews).  First observed in the SU Ursae Majoris stars, {\it common}
(a.k.a. {\it positive} or {\it apsidal}) superhumps have periods a few
percent in excess of their system orbital periods.  The story and
classification scheme, however, have in recent years become more
complicated.  First, some old novae and nova-likes show ``permanent''
positive superhumps indicating a high and roughly constant mass
transfer rate, e.g., V603 Aquilae (Patterson et al.\  1993, 1997) V795
Herculis (Patterson \& Skillman 1994), H 0551-819 (Patterson 1995), MV
Lyrae (Skillman, Patterson, \& Thorstensen 1995), and V1974 Cygni
(Retter, Leibowitz, \& Ofek 1997; Skillman et al.\ 1997).  Second, a
few systems, including V603 Aql, H 0551-819, V503 Cyg
(Harvey et al. 1995), V1159 Orionis (Patterson et al.\ 1995), TT
Arietis (Skillman et al.\  1998; Kraicheva et al.\ 1999), AM CVn
(Skillman et al.\ 1999) display so-called {\it negative} (or nodal)
superhumps, which have periods a few percent {\it shorter} than the
orbital period.  

The favored model explaining positive superhumps is that for systems
in the range $0.03 \lesssim q \lesssim 0.33$ with high mass transfer
rates and viscosities, the outer disk can expand to the radius near the
3:1 eccentric inner Lindblad resonance (just outside the 3:1
co-rotation radius) where it can be driven periodically by the tidal
field of the orbiting secondary (Whitehurst 1988, 1994; Hirose \&
Osaki 1990; Lubow 1991a,b; Murray 1996).   Numerical simulations by
Simpson \& Wood (1998) show that over one oscillation period,
as viewed face-on, the disk is pulled into a distorted, elongated shape
followed by relaxation back to a nearly circular shape, and that
during this oscillation, the spiral shock pattern advances $180^\circ$
in the slowly precessing frame (n.b., spectroscopic time-series of an
eclipsing superhumper may reveal this motion).  Simulations further
suggest that viscous energy dissipation through the cycle is the dominant
source of the observed photometric oscillations and is highest when
the disk is maximally distorted and the streamlines are most
non-circular.

Because both modes have been observed {\it simultaneously} in several
of the systems listed above (e.g., V603 Aql, V1159 Ori, H 0551-819,
and V503 Cyg), the physical origin of the negative superhumps must be
distinct from the above mechanism.  The model suggested by Harvey et
al.\  (1995) invokes the retrograde precession of the {\it line of
nodes} in an accretion disk {\it tilted} with respect to the orbital
plane (see also Bonnet-Bidaud, Motch, \& Mouchet 1985 and
Barrett, O'Donoghue, \& Warner 1988).  Having identified physical
processes associated with each superhump mode, we hereafter follow
Harvey et al.\ (1995) and refer to positive and negative superhump
modes as apsidal and nodal, respectively, thereby avoiding the awkward
term ``negative period excess.''

The linear theory of tilted disks (Papaloizou \& Pringle 1983;
Papaloizou \& Terquem 1995) suggests that for the disk to precess as a
solid body, the sound crossing timescale must be shorter than the
precession timescale, a condition that is well satisfied in our
simulations.  
Larwood
et al.\ (1996) studied the general dynamics of tilted disks using the
method of smoothed particle hydrodynamics (SPH), and found that for a
constant surface density $\Sigma$ disk tilted at angle $\delta$ with
respect to the orbital plane, the nodal precessional rate is given by
\begin{equation}
{\omega_p\over\Omega(R)} = -{15\over32}q\beta^3\cos\delta
\label{eq: prec}
\end{equation}
where $\beta\equiv R_d/a$ is the disk radius in units of the orbital
separation, and where a ratio of specific heats $\gamma=5/3$ was
assumed.  For typical values, $q=0.25$, $\beta=0.4$,
$\Omega(R)=3\Omega_{\rm orb}$, equation \ref{eq: prec} gives $\sim$40
orbits per precession period, which is in factor-of-two agreement with
observations and our simulations below.

In this work we report the preliminary results of a large parametric
study (Montgomery, Wood, \& Simpson 2000, in preparation) exploring
the dynamics of tilted disks and the origin of the nodal superhump
phenomenon.  Our primary result is that the model of a tidally-driven
retrogradely-precessing disk is basically confirmed, but that the
driving frequency is slightly higher than twice per orbit as each of
the two out-of-orbital-plane halves of the disk are perturbed
sequentially and out of phase with each other.   Because we receive
photons from only one face of an optically-thick disk in
superoutburst, we {\it measure} a photometric period which is just
shorter than the orbital period.  This model predicts that the
waveform observed from a tilted disk that is optically-thin or viewed
from high inclination should be approximately doubly-periodic each
orbit.

\section{General Characteristics of the SPH Models}

Smoothed particle hydrodynamics (SPH) provides an effective tool for
modeling accretion disks in general, and those arising in binary star
systems in particular (for a review, see Monaghan 1992).   We use the
code described in Simpson (1995) and Simpson \& Wood (1998).  The code
is based on particle-particle interactions within the disk and the
mass transfer stream, with the only remaining body forces being the
gravitational attraction of the two stars (i.e., disk self-gravity is
negligible and ignored).  In the models reported here, particles
are injected into the disk through the binary's inner Lagrangian
($L_1$) point at a rate of 2,000 particles per orbit until the maximum
of 25,000 particles is reached.  The number of particles is kept
constant after this point by inserting a new particle
whenever an already existing particle leaves the system by
accretion or ejection from the computational space.  We note that
once equilibrium is established, the typical particle injection rate through 
$L_1$ is a nearly-constant 280 particles per orbit (1.4 per system
timestep), giving a mean particle lifetime of 89.4 orbits from
injection to accretion.  At this point the disk is at the same
equilibrium as if we had {\it set} the injection rate to a constant 280
particles per orbit.

The fundamental timestep for the simulations is $P_{\rm orb}/200$, and
individual particles can have shorter timesteps if needed, in steps of
factors of 2.  All particles in the simulation have identical
smoothing lengths $h$ for computational efficiency.  We note that
although our constant-$h$ disks are somewhat under-resolved in the
inner disk region, the superhump oscillations are located in the outer
disk and are well-resolved. 

For the runs presented here we assume an ideal gamma-law equation of
state $P=(\gamma-1)\rho u$, where $u$ is the internal energy and where
we use $\gamma=1.01$.  The artificial viscosity prescription used is
that of Lattanzio et al.\ (1986) with $\alpha=0.5$, $\beta=0.5$, and
$\eta=0.1h$, as recommended by Lombardi et al.\ (1999).

As in Simpson \& Wood (1998), we calculate an energy production 
time series --- effectively a bolometric light curve ---
by assuming that the instantaneous disk luminosity is
proportional to the change in the summed internal energy change of all
particles integrated over the previous system timestep.  
The internal energy change is typically dominated by viscous energy
dissipation, but $P\,dV$ work also contributes.  We use Fourier
analysis to determine the frequencies at which the disk is driven by
the tidal field of the secondary.

\section{Results}

\subsection{The Reference Simulation Run}

In this work we report the results for a $q=0.25$ system with assumed
masses $M_1=0.8M_\odot$ and $M_2=0.2M_\odot$.  With our choice of
secondary mass-radius relationship $R_2 =
(M_2/M_\odot)^{13/15}R_\odot$ (Warner 1995, \S2.8.3) and the relation
for the volume radius of the secondary's Roche lobe (Eggleton 1983)
  \begin{equation}
{R_2\over a} = {0.49q^{2/3}\over 0.6q^{2/3}+\ln(1+q^{1/3})},
\end{equation}
our simulation unit length scales to $a=0.93R_\odot$ and 
period $2\pi$ scales to $P_{\rm orb}=2.47$~hr.
We note that our results below for 
the disk oscillation and precession frequencies in units
of $\orb^{-1}$ would be unchanged for a simulation with
$M_1=0.6M_\odot$, $M_2=0.15M_\odot$, and $P_{\rm orb}=2.0$ hr --- the
reader should place no special significance on the period here falling
in the CV period gap. 

\begin{figure} 
\plotone{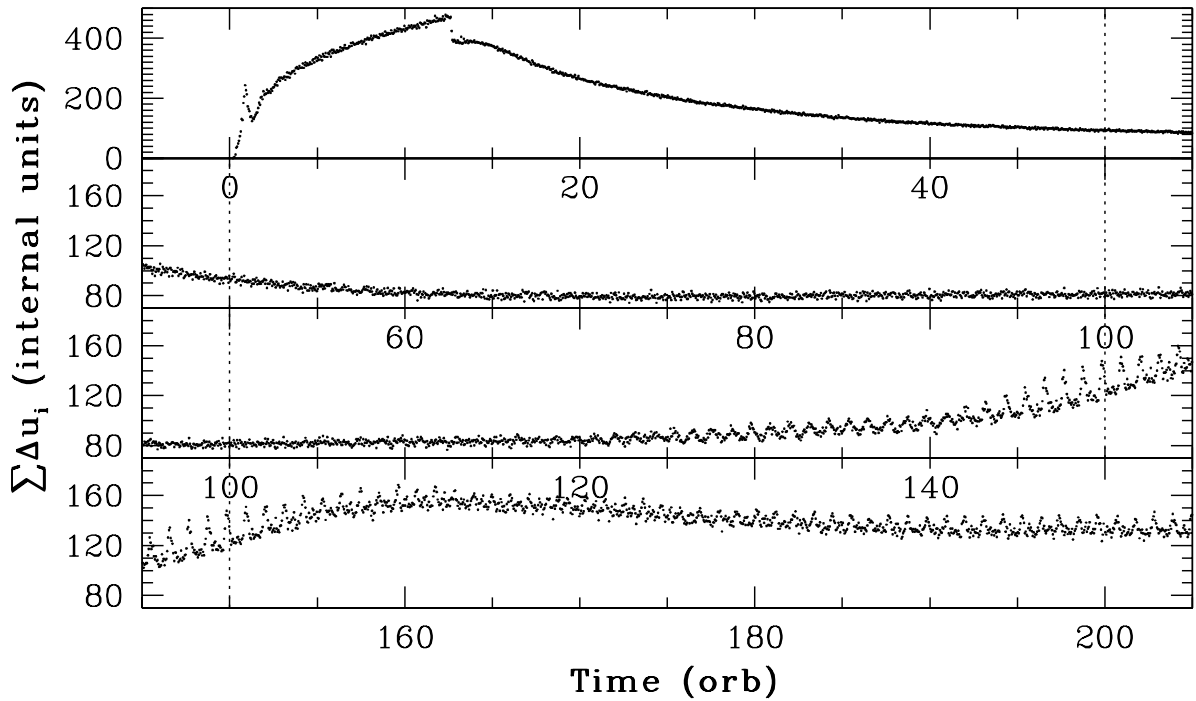}
\caption{We show 200 orbits of the simulation light curve for a
$q=0.25$ system.
Common (positive) superhumps are evident starting near orbit 110 and
reach a dynamical equilibrium state near orbit 200. Note the vertical
scale spans 500 units in the top panel and 100 units in the bottom 3
panels. Five orbits are repeated between panels for clarity.}
 \label{fig: lc200}
\end{figure}

Figure~\ref{fig: lc200} shows the simulation light curve of the
$q=0.25$ apsidal superhump oscillation out to orbit 250, at which
point the waveform is essentially stationary.  For the purposes
display only, we sum 10 points to suppress the high-frequency noise
but the ``data'' are otherwise unprocessed.  Each point is thus a
relative integral average over an interval $P_{\rm orb}/20$. 

The apsidal superhumps reach observable amplitude near orbit 110.  As
the oscillation grows the global dissipation rate and mass flux
through the disk rises, increasing the overall disk energy production
(luminosity).  The system is eventually forced into a state of
dynamical equilibrium near orbit 200.  The artificial light curves of
the apsidal superhumps are double humped and reminiscent of published
lightcurves.

The Fourier amplitude spectrum of orbits 200 to 300 is shown in the
top panel of Figure~\ref{fig: FTs}.  The double-humped profile yields
a second harmonic of amplitude comparable to that of the fundamental
frequency $\nu_a$.  The third through fifth harmonics are also clearly
present with decreasing amplitudes.  The frequencies of the peaks in
this Figure are listed in Table~\ref{tbl: FT peaks}.

\subsection{Dynamics of a Tilted Disk}

\begin{figure} 
\plotone{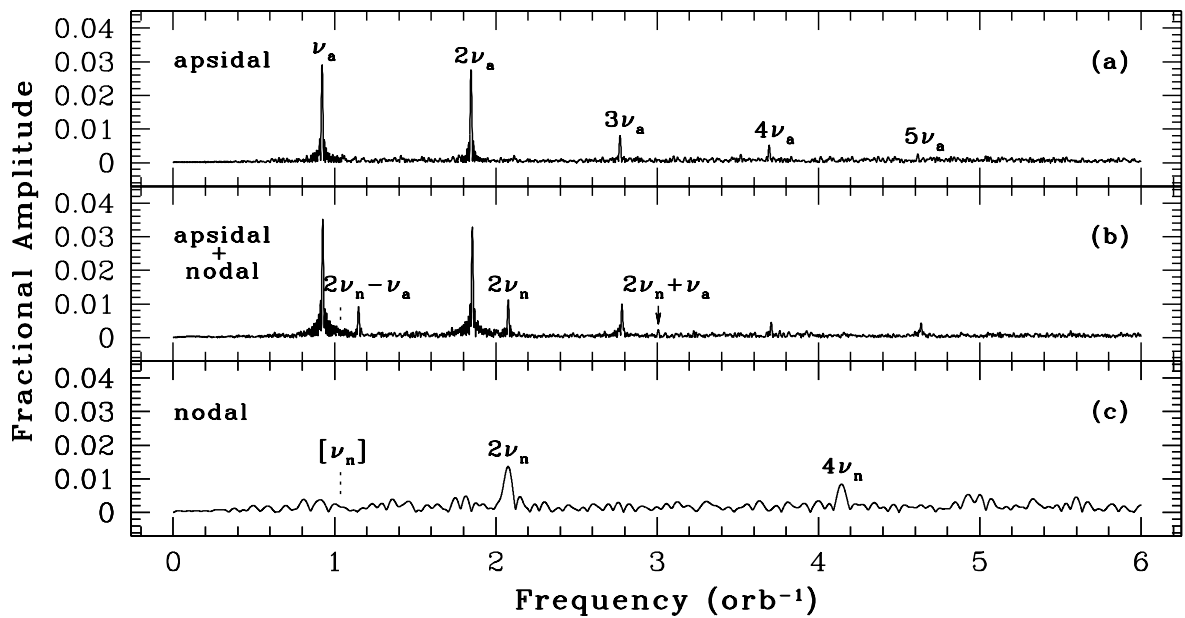}
\caption{The top panel shows the Fourier amplitude spectrum 
of orbits 200 to 300 of the untilted reference disk. 
The middle panel  shows the amplitude spectrum of orbits 200 to
300 of a disk tilted by $5^\circ$ at orbit 200.  In addition to the
apsidal frequencies, there is now power at a peak labeled $2\nu_n$, 
and at the linear
combination frequencies
$2\nu_n - \nu_a$ and $2\nu_n + \nu_a$ (labels are centered over the
peaks in all cases). 
The bottom panel shows the amplitude spectrum of orbits 50 to
70 of a disk tilted at orbit 40, well before apsidal superhumps
develop.  The driving frequency $2\nu_n$ and
its harmonic are clearly evident.  Observed nodal
superhumps have frequencies $\nu_n$ one-half this frequency (dotted
line), as we see
only one side of an optically-thick disk in superoutburst.
}
 \label{fig: FTs}
\end{figure}

\begin{figure} 
\plotone{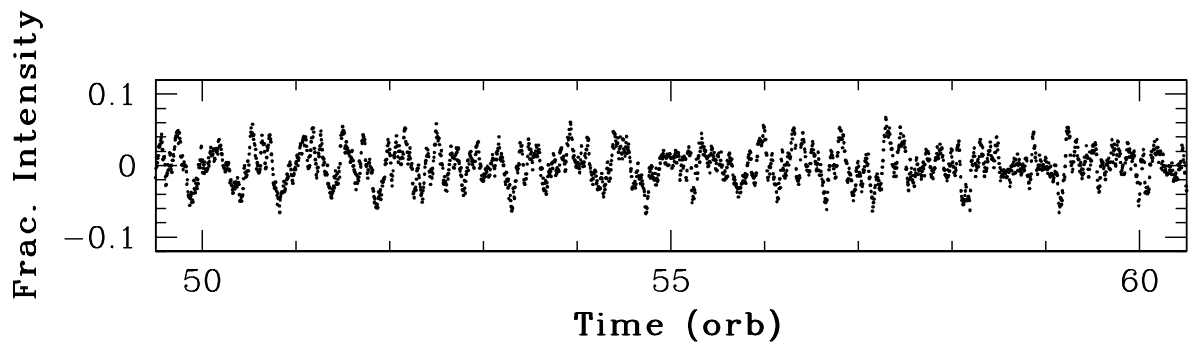}
\caption{Boxcar-smoothed light curve showing nodal superhump
oscillations (filter width is 11 points).}
 \label{fig: lc.50-60}
\end{figure}

To test the idea that the origin of negative (nodal) superhumps is the
periodic tidal stressing of a tilted disk with a regressing line of
nodes, we first artificially rotate the position and velocity vectors
of all particles 
by $5^\circ$ at the end of orbit 200, where the rotation is about the line
perpendicular to the line of stellar centers and passing through the primary.
Running the simulation again to orbit 300, we obtained the Fourier
transform in the middle panel of Figure~\ref{fig: FTs}.  We thought at
first the power at just above the orbital frequency was the nodal
superhump frequency, but did not understand the physical origin of the
other new peaks.  We then tilted the disk at orbit 40 (well before the
onset of the apsidal mode) allowed 10 orbits for transient effects to
die away, 

We show the energy-production curve from the second run for orbits 50
to 60 in Figure~\ref{fig: lc.50-60}, where for display purposes, we
have boxcar-smoothed the data with a filter of width 11 points
($\sim$0.05 orb).  The troughs are easy to identify in the noisy data,
and there is an oscillation evident with a period of roughly
$\sim$$P_\orb /2$.  We calculated the Fourier transform of this
energy-production curve from orbits 50 to 70 (bottom panel of
Figure~\ref{fig: FTs}), stopping before the apsidal mode appears.
{\it Only} the frequency with $P\sim P_\orb /2$ and its second harmonic
are detected. 

At this point the underlying physics is clear: for an axisymmetric
disk tilted about the primary, once per orbit the tidal field disturbs
in turn the fluid flow in {\it each of the two} disk halves which are
out of the orbital midplane and separated by the line of nodes.  Each
tidal disturbance is asymmetrical with respect to the disk midplane in
that it affects the side facing the orbital midplane more than the
opposite side of the disk.  For an optically-thick disk in
superoutburst, only photons from one side of the disk are observed,
and so the observed nodal superhump periods are just shorter than once
per orbit as the tilted disk precesses in a direction opposite the the
fluid flow in the disk.  To maintain consistency with the
observational literature, we have labeled the main peak in the bottom
transform $2\nu_n$.  The location of what would be the observed nodal
superhump frequency is marked with a dotted line, and is included in
brackets in Table 1.

\begin{figure} 
\plotone{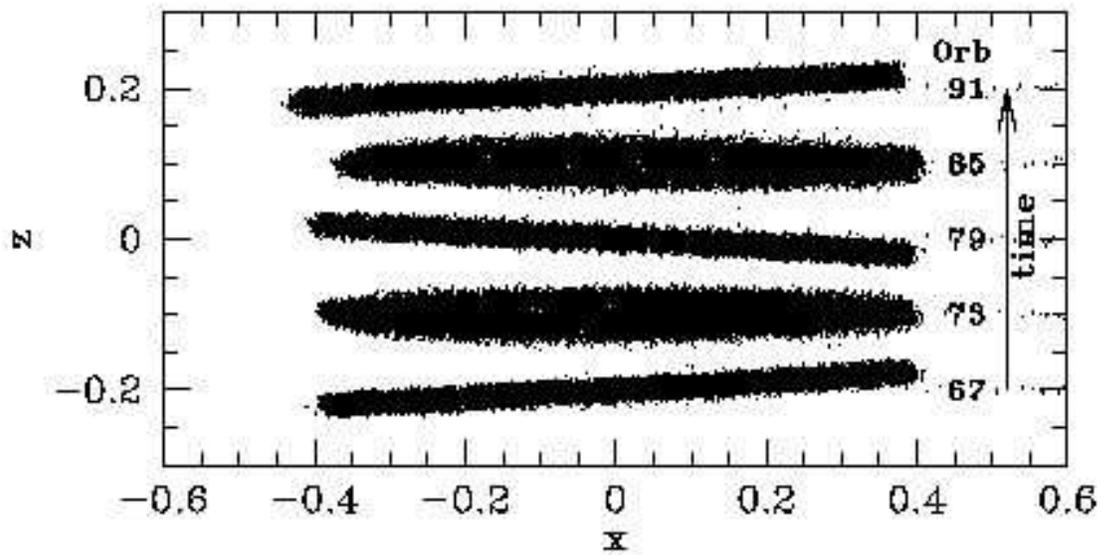}
\caption{Sideview snapshots spanning one full nodal precession cycle
of 24 orbits. The disk fluid has an angular momentum vector with a positive $z$
component, but the tilted disk precesses in the opposite sense.}
 \label{fig: sideview}
\end{figure}

Figure \ref{fig: sideview} shows 5 sideview snapshots over a full
nodal-precession period of $P_{np}\approx24$ orbits for the $q=0.25$
disk.  The wobble is obvious.  Because mass is added at the disk
midplane, the tilt must decay with time.  To quantify the decay time
scale, we ran this simulation out to orbit 400, and computed the
Fourier amplitude spectra of orbits 250-300, 300-350 and 350-400.
Assuming an exponential decay of the form $A(t_2)=A(t_1) e^{-\Delta
t/\tau_d}$ with times at the centers of the above intervals, we use
the amplitudes of the $2\nu_n$ peak in the three spectra ($A =
0.0104$, 0.00714, and 0.00492, respectively) to derive a decay time
scale $\tau_d= 134$ orb. The corresponding ``half-life'' of the tilt
decay, $\tau_{1/2}=\ln 2\cdot \tau_d=93$ orb, is very nearly the
mean particle lifetime, as it must be from conservation of angular
momentum arguments.

Now we can fully understand the Fourier transform of the reference
disk tilted at orbit 200 and spanning orbits 200 to 300
(Figure~\ref{fig: FTs}, middle panel).  First, tilting the disk did
not significantly affect the frequencies of the apsidal superhump
modes.  Second, the power at the frequencies $2\nu_n-\nu_a$ and
$2\nu_n+\nu_a$ (see Table 1) are linear combination frequencies
resulting from the non-linear response of viscous dissipation in a
disk which is precessing in opposite directions simultaneously and
suffering tidal driving from the orbit of the secondary.  We note that
while the $2\nu_n+\nu_a$ peak is small in this transform, it is quite
pronounced in disks with larger tilts.

\subsection{On Generating a Tilted Disk}

Lubow (1992) explored the growth rate of the tilt instability mode
and found it to be $\sim$1/50-th that of the apsidal superhump mode
growth rate.  Murray \& Armitage (1999) confirmed  a
consistent-with-zero growth rate mode using three-dimensional smoothed
particle hydrodynamics (SPH) simulations where particles were injected
into the disk at the circularization radius after an initial burst of
particles to build the disk rapidly.  

Several models have been suggested to generate tilted disks in
cataclysmic variables, including an $L_1$ point effectively displaced
out of the orbital plane as a result of channeling of the accretion
flow near $L_1$ by magnetic activity associated with the mass-losing
secondary (see, e.g., Barrett et al.  1988).  As a simple test of this
idea, we ran a simulation with the same parameters as our 
reference run above and another with $q=0.05$, but with the $L_1$
injection point displaced in $z$ by $0.05a$.  The accretion stream
passes through the disk midplane and impacts the disk on the side
opposite the displaced $L_1$ point, and at a radius of about one-third
the disk radius.  Because the stream does not impact the disk rim, the
disk radius grows faster than in the reference simulation run, and
common superhumps onset at about orbit 50 versus orbit 110 for the
reference run.  However, {\it we find no significant power at the
frequency $2\nu_n$} in the amplitude spectrum of either simulation run
as a result of displacing the $L_1$ point in a simulation run for 100
orbits. 

If local magnetic field evolution and recombination contributes to the
fluid viscosity and disk energetics, it is quite plausible that an
outburst event could be asymmetical with respect to the disk midplane
and hence result in a tilted disk.  Our code is not sufficient to
explore this idea, however, so for now the origin of the disk tilt
remains a puzzle.

\section{Summary}

Although these are preliminary results, it appears the
tilted-disk model neatly explains many of the observational
characteristics of nodal (``negative'') superhumps, and also suggests
further observational and numerical tests.  Our findings here can be
summarized as follows.

\begin{itemize}

\item The results are consistent with the nodal regression model for
the origin of negative superhumps.  We tilted a disk both before and
after the development of common apsidal superhump oscillations, and
find power at twice the expected (observed) nodal superhump frequency.
In superoutburst the disk is optically thick and so we observe only
one face and hence only a single nodal mode brightening event per
orbit --- the other occurs on the opposite (hidden) face of the disk
and is $180^\circ$ out of phase.  This model predicts that optically
thin and/or edge-on tilted disks should display power at $2\nu_n$.  

\item In disks oscillating in both nodal and apsidal modes
simultaneously, there is also power at the combination frequencies
$2\nu_n-\nu_a$ and $2\nu_n+\nu_a$.  These combination modes have not
yet been identified in nature, but if demonstrated would provide
additional support for this model. 

\item In our simulations, where we keep the particle number constant,
the amplitude of the nodal mode decreases on a timescale of $\sim$1000
orbits (several weeks if scaled to a physical orbital period).  This
is consistent with the observations, where the nodal superhump modes appear
to have a much longer decay time than the apsidal modes.

\item The origin of the disk tilt remains an open question. Simply
displacing the $L_1$ point out of the orbial midplane does not appear
to be sufficient to generate a tilted disk, but more simulations are
required. 

\end{itemize}

\acknowledgments
We thank Fred Ringwald, Joe Patterson, Darragh O'Donoghue, and Shannon
Baker-Branstetter for useful discussions.  We also thank the anonymous
referee.  This work was supported in part by NASA through grant NAG
5-3103.

\clearpage


\begin{deluxetable}{ccc} 
\tablewidth{4in}
\tablecaption{Apsidal and Nodal Mode Frequencies}
\tablehead{
\colhead{Mode} & \colhead{Frequency\tablenotemark{a}}     & \colhead{Fractional}\\
               & \colhead{(orb$^{-1})$} & \colhead{Amplitude}
}
\startdata
$\nu_a$ & 0.92 & 0.030 \\
2$\nu_a$ & 1.85 & 0.028 \\
3$\nu_a$ & 2.77 & 0.010 \\
4$\nu_a$ & 3.69 & 0.0052 \\
\vspace{3\jot}
5$\nu_a$ & 4.64 & 0.0025 \\
$[\nu_n]$ & [1.04] & $\cdots$ \\
$2\nu_n$ & 2.08 & 0.011\\
\vspace{3\jot}
$4\nu_n$ & 4.14 & 0.0084\\
$2\nu_n-\nu_a$ & 1.15 & 0.0092\\
$2\nu_n+\nu_a$ & 3.01 & 0.0002\\
\enddata
\tablenotetext{a}{The uncertainties on all frequencies except $4\nu_n$ are
$0.01 \rm\ orb^{-1}$ (100-orbit transform); uncertainty on $4\nu_n$ is 
$0.05 \rm\ orb^{-1}$ (20-orbit transform).}
\tablenotetext{b}{Not actually present in the Fourier transform of our
simulation data, this is the nodal superhump frequency that would be 
observed (see text).}
\label{tbl: FT peaks}
\end{deluxetable}


\clearpage
\onecolumn


\end{document}